# Towards the Structure and Mechanisms of Complex Systems, the Approach of the Quantitative Theory of Meaning

Inga Ivanova[1] and John S. Torday[2]

## Abstract

We study analysis of complex systems using a Quantitative Theory of Meaning developed as an extension of Shannon's Communication Theory. The approach considers complexity not in terms of the manifestation of its effects which are manifestation of the dynamics of the system, but in terms of primary causes and taking into account the topology of the system. Here, the dynamics of the system are provided by reflexive communication between heterogeneous agents that make up the system. Unlike Shannon's Communication Theory the Theory of Meaning imposes restrictions on the complex systems being analyzed. Non-linearity and specific dynamics of the system arise as a consequence of the topology of the system. This topology also suggests a method for analyzing complex systems, the logistic Continuous Wavelet Transform (CWT). The paper also lays the foundation for future research in various fields studying complex systems of interacting heterogeneous agents, which may form a new paradigm for better understanding the structure, mechanisms, and dynamics of complex systems.

**Keywords:** Complexity, information, meaning, Triple Helix, model.

---

[1] corresponding author; Institute for Statistical Studies and Economics of Knowledge, National Research University Higher School of Economics (HSE University), 20 Myasnitskaya St., Moscow, 101000, Russia inga.iva@mail.ru;

[2] Professor of Pediatrics Obstetrics and Gynecology Evolutionary Medicine, UCLA, jtorday@ucla.edu

## 1. Introduction

Complex systems play an important role in the natural world around us. A complex system can be defined as a set of components that interact with each other in such a way that the behavior of the system as a whole is not simply the sum of the behaviors of its parts (Bar-Yam, 2004).

Complex systems are typically characterized by nonlinearity, feedback loops, and emergent behavior. The essential attributes of complex systems are: the presence of different dynamics that interact with each other, self-generation, ramified communication networks, the perception of reality from different perspectives, and giving meaning to the perceived information from these different perspectives, which determine the system's response to a changing environment (Trigona and Cianci, 2014). An important property of complex systems is nonlinearity, which means that the total effect of the interaction of different parts of the system produces an effect that is not equal to the sum of the effects. This effect can be either greater than the sum of the effects in the case of synergy or less than the sum of the effects in the case of dissonance.

The study of complex systems, which consist of numerous interacting components whose collective behavior is often unpredictable, increasingly relies on the use of computers. But this approach requires a large number of calculations. In addition, the results obtained do not always allow us to identify the mechanisms that determine the dynamics of the systems. Another way to study complex systems is information theory, for example, Shannon's Communication Theory. The application of Shannon's Communication Theory (Shannon, 1948) to complex systems focuses on understanding how information is transmitted and processed within the system,

The cornerstone of Shannon's Communication Theory—quantification of information using entropy—offers valuable insights for the analysis of complex systems. The theory provides a framework for analyzing how information is encoded, transmitted, and decoded through various channels, emphasizing the importance of noise and redundancy in communication. The theory provides a robust foundation for analyzing complex systems, offering tools for quantifying information flow, measuring complexity, and understanding interactions between components. Despite some criticism regarding its simplifications, the principles derived from Shannon's work remain relevant in fields as diverse as ecology, social science, and engineering.

The concept of redundancy, a key aspect of Shannon's theory, plays an important role in the robustness of complex systems. Redundancy in this context is the presence of multiple pathways or mechanisms that perform similar functions. This redundancy ensures that the system is resilient to component failures. For example, in social networks, redundancy of connections ensures that information can still be transmitted even if some nodes or links fail (Albert *et al*., 2000). Similar redundancy is observed in biological systems, such as multiple pathways for metabolic processes that ensure survival even in the face of genetic or environmental perturbations. Analyzing redundancy in complex systems using information-theoretical measures such as correlation between information channels provides insight into the robustness of the system to perturbations.

Another concept is mutual information, which quantifies the statistical dependence between two or more variables in a system, offering insight into the relationships and interactions between different components (Cover & Thomas, 2006). This measure has been used to analyze gene regulatory networks (Steuer *et al*., 2002) and to understand the flow of information in complex brain networks (Sporns, 2010). Similarly, transfer entropy measures the directional flow of information, thereby providing a more nuanced understanding of causal relationships in a system. Using mutual information, researchers can quantify the dependencies between different components of a system, analyzing interactions in complex systems.

A significant application of Shannon's theory in complex systems is the use of information measures as indicators of complexity. The amount of information needed to describe a system correlates with its complexity. This theory has proven to be highly applicable to the study of complex systems where information generation and processing are critical. For example, in biological systems, information entropy has been used to quantify genetic diversity (Li & Vitányi, 2008), biodiversity in a population (Monleón-Getino, Rodríguez-Casado, & Verde, 2019), and the analysis of neural activity (Strong *et al*., 1998). Higher entropy in these contexts often indicates greater complexity and adaptability.

Despite its significant contributions, the application of Shannon's theory to complex systems also faces limitations. 1) Shannon's framework quantifies information in terms of uncertainty reduction. However, in biological or social systems, the semantic content of information, its meaning, and its functional role are critical. This distinction between information and meaning is the main difference between his theory and traditional communication systems, such as social communication. Shannon's theory simplifies communication by treating it as a linear process devoid of context or meaning. 2) In reality, communication often involves feedback loops and dynamic interactions that are not captured by traditional models. Many complex systems exhibit non-stationary behavior, with their statistical properties changing over time. A simple measure of entropy does not fully capture the dynamic interactions of information flow within the system. In addition, the application of Shannon's framework requires careful consideration of the specific characteristics of complex systems, including their non-linear dynamics and adaptive behavior. The inherent limitations of the classic Shannon framework, especially with respect to non-stationarity and purely syntactic definition of information, require the development of more advanced methods that explicitly take into account the unique characteristics of complex systems.

In this paper, we show how a Quantitative Theory of Meaning, developed as an extension of Shannon's Communication Theory, can be applied to the analysis of complex systems of different origins. Starting from the problem of configurational information sing alteration in three or more dimensions (e.g., McGill, 1954; Yeung, 2008, p. 59f.) and modifying Shannon's general communication scheme in terms of the positional differentiation of communicating agents (Leydesdorff and Ivanova, 2014), we link this framework to the Triple Helix (TH) model (Etzkowitz and Leydesdorff, 1995, 1998). TH model, originally developed as a model of university-industry-government relations, is the simplest example of an inherently nonlinear self-organizing system that can serve as a building block for complex systems of higher dimensions. Thus, the exchange of information in a complex system and the subsequent dynamics of the system are provided by the structure of these systems. Future research should continue to adapt and expand these fundamental concepts to better capture the subtleties of complex systems structure, dynamics and mechanisms.

## 2. Information and meaning in communication system

From an information exchange perspective, complex systems can be viewed as information ecosystems. Information ecosystem theory is built on the concepts of ecosystem and information. The concept of ecosystem originated in ecological studies. From an ecological perspective, an ecosystem can be defined as "*recycling flows of nutrients along pathways made up of living*

*subsystems which are organized into process-orientated roles; connects living and non-living subsystems*" (Shaw and Allen, 2018, p. 90). There are three points to note in this definition: "*living subsystems*", "*nutrient flows*", and "*process-oriented roles*". More broadly, in other domains, this can be described as "*agents*", "*interaction*", and "*heterogeneity*". In terms of social systems, Kühn (2023) proposed a definition of an information ecosystem as "*all structures, entities, and agents related to the flow of semantic information relevant to a research domain, as well as the information itself*". Thus, an ecosystem implies a nonlinear self-organizing system consisting of a set of agents, where the system dynamics are the result of communication between agents, where the transmitted information is essentially semantic. That is, in the sender-receiver communication scheme, information, as a "message", is provided with meaning on the sender's side, and the meaning is extracted from the message on the receiver's side. In the case where the sender and receiver consider the same message from different points of view, these can be two different meanings attached to the same message. Thus, communication can generate an excess of meanings.

Leydesdorff and Ivanova (2014), building on Shannon's (1948) information theory, proposed a way to quantify semantic communications using mutual redundancy as a measure of the surplus of meanings that can generated in reflexive communications. The origin of the concept of quantifying meaning can be traced back to the TH model of innovations. The development of the Triple Helix model of university-industry-government relations as a concept of a national innovation system (Etzkowitz & Leydesdorff, 1995, 1998) required a method for assessing the effectiveness of interactions between actors that would help assess the extent to which a system comprising actors interacts in a coordinated (or synergistic) manner. Leydesdorff proposed using mutual information in three dimensions to assess the synergy of a system. Venn diagram of overlapping uncertainties in two variables is presented in Fig. 1

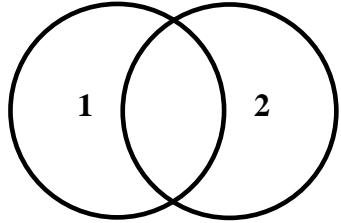

**Figure 1**: Venn diagram of overlapping uncertainties in two variables.

Shannon defined mutual information $T_{12}$ as the central overlap of two uncertainties: $T_{12} = H_1 + H_2 - H_{12}$. Here uncertainties $H_i$ are defined in terms of probabilities $p_i$ with respect to possible states of the system: $H_i = -p_i log p_i$. $H_{12}$ – is the resulting uncertainty after the overlap. Since the surface area decreases after overlap - $T_{12}$ is always positive. When a third variable is added (Fig. 2), the central overlapping area represents configurational information (e.g., Abramson, 1963), which is no longer Shannon's mutual information because it can be negative.

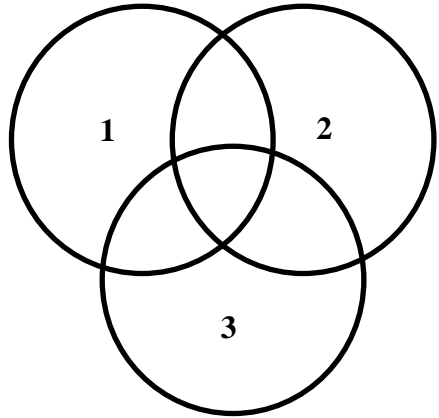

**Figure 2**: Venn diagram of overlapping uncertainties in three variables 1, 2, and 3.

Adding higher dimensions successively changes the sign of the configurational information from plus for even dimensions to minus for odd dimensions. The sign change problem can be solved by considering a measure of "excess " information - $R_{12}$. Instead of subtracting overlapping areas one adds to these areas. E.g. for two dimensions:

$$H_{12} = H_1 + H_2 + R_{12} \qquad (1)$$

The resulting measure can be conceptualized as redundancy. The following relationships between redundancy and configuration information hold:

$$R_{12} = -T_{12}$$

$$R_{123} = T_{123} \tag{2}$$

$$\dots$$

$$R_{12\dots n} = (-1)^{n-1}T_{12\dots n}$$

This explains the alternation of signs. Negative redundancy indicates a decrease in uncertainty. Leydesdorff conceptualized the mechanism of increasing uncertainty in overlap in terms of an excess of additional, but not yet realized, options generated during reflexive information processing (Leydesdorff, 2010). This required modification of Shannon's general communication system by adding additional B and C levels (Fig. 3).

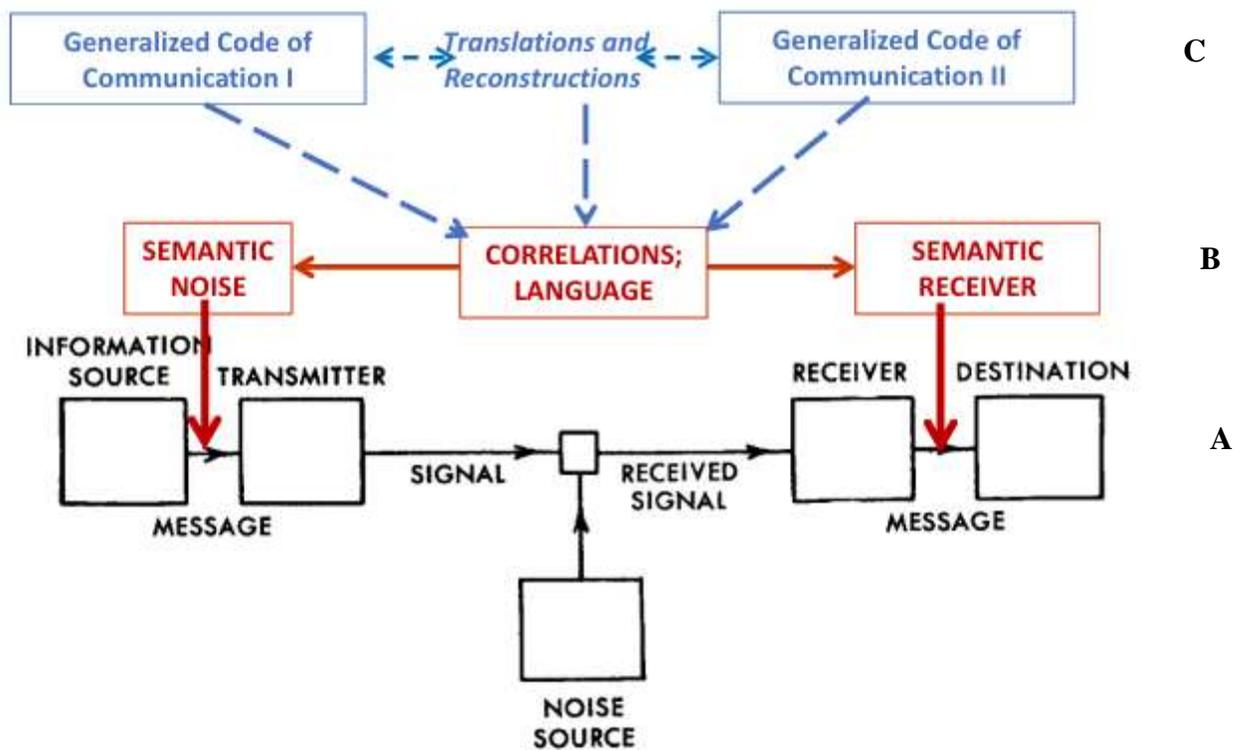

**Figure 3**: Shannon diagram of general communication system (level A) with added levels B and C (From Leydesdorff, Petersen, and Ivanova, 2017)

Information transmitted from the sender to the receiver at level A is subject to semantic encoding and decoding on the sender and receiver sides (blocks labeled "semantic noise" and "semantic receiver", which encode and decode the meaning into messages). The encoding and decoding of information is performed using appropriate sets of communication codes, which are subject to translation and reconstruction during the communication process. The sets of communication codes may differ for different senders and receivers and partially overlap (i.e., some codes may coincide, and some may not). The codes can be represented as eigenvectors in a communication matrix that span orthogonally and introduce functional differentiation between the codes. Thus, communicating agents can be considered as "positionally differentiated". Semantic coding gives meaning to information. When sets of communication codes do not coincide, the same information can be provided with different meanings in the sender and receiver. The meaning of the information can be used to generate options of future development. Communication between positionally differentiated agents that make up the system leads to proliferation of (not yet realized) options of possible future development of the system. Thus, the number of actual and possible states of the system increases, which is captured by the "excess" information measure, which increases the maximum possible uncertainty.

Reflexive communications generate a complex network structure. While communications include a network structure of relations, there is another network of correlations on the top of it. The network of relations and network of correlations can be viewed as local and non-local networks which generate local and non-local interactions. Communications are taking place recursively - with the arrow of time, and options are generated with respect to the future, i.e. against the arrow of time. The combination of these processes generates loops of feedforward and feedback communications.

The introduction and conceptualization of the measure of "excess" information marked the transition from TH model of university-industry-government relations as a model for studying innovation to the conceptualization of a Quantitative Theory of Meaning.

Another notable feature of the TH model is related to its self-organization. TH model can be conceptualized in terms of its components – institutional spheres such as university, industry and government, relationships – connections between institutional spheres and functions – novelty production, wealth generation, and normative control (Etzkowitz and Ranga, 2012). In terms of the relative influence of institutional spheres, three modes can be distinguished: TH I, where government drives science and industry, TH II, with a dominant role of markets and TH III, where science has a predominant importance (Etzkowitz and Leydesdorff, 2000). Since the TH system is in constant transition, the modes successively replace each other. The process can be depicted as a vector representing the TH system rotating in a three-dimensional Cartesian coordinate system whose axes represent TH actors. Rotations in the three-dimensional coordinate system are non-commutative, i.e. the order of rotations cannot be changed without changing the outcome. This fact, combined with the assumption that the innovation process is described in terms of innovation waves from a purely formal mathematical point of view, leads to an equation describing institutional actors interacting through a nonlinear communication field (Ivanova and Leydesdorff, 2014). The nonlinearity of the field leads to self-interaction, so that not only do actors generate communications, but communications also generate communications. This entails a self-replication of the TH system, leading to a fractal structure of the system. This replication is driven by path dependencies in the order of operation among the institutional actors representing the three selection environments. In this respect, the TH system is inherently a nonlinear system, in contrast to the Double Helix system, which can be seen as linear[3]. The fractal structure of TH leads to replication of the system on smaller scales. Accordingly, subsequent innovation systems at smaller scales – such as sectoral and regional innovation systems – have a topology similar to the national innovation system in which they are embedded. This similarity is observed in the EU research domain. While the entire EU research domain is scale-invariant, smaller research networks are assumed to have the same topology. Bigiero and

---

[3] Since the order of successive rotations in a two-dimensional coordinate system can be swapped without changing the result.

Angelini (2015) show that some R&D collaboration networks in the European aerospace research domain have a scale-free topology. A scale-free topology implies a fractal structure that is generated through self-organization (autocatalysis) and increases the resilience of the system.

### 3. Operationalization

Assuming that the number of generated variants is proportional to the intensity of communications, the redundancy R generated in the system can be represented as the sum of positive and negative terms:

$$R(t) = P^2(t) - Q^2(t) \qquad (3)$$

The positive term indicates an increase in entropy, and the second term destroys the existing entropy. According to the second law, entropy in non-equilibrium systems always increases with time, so the first term $P^2$ can be associated with the historical development of the system revealed in the implementations. The reduction of uncertainty can be achieved through self-organizing processes, and therefore the term $Q^2$ can be attributed to the synergy that is generated within the system during its evolution.

It is interesting to note that Eq. 3 resembles the expression for enigmatic information (James, Ellison, and Crutchfield, 2011)

$$Q[X_N] = T[X_N] - B[X_N] \qquad (4)$$

Here $X$ is the distribution of a random variable, $N$ is the number of distributions, $Q[X]$ is equal to the multidimensional mutual information for $N$=3, which is given by the redundancy $T[X_N]$ minus the synergy $B[X_N]$ and can be negative.

The process of autocatalysis can be illustrated using the principle of "triadic closure" (e.g., Sun & Negishi, 2010, Bianconi *et al*., 2014). When the interaction between two agents is disrupted by a third, an autocatalytic cycle can be formed. Two modes can be distinguished: self-organizing (Fig. 4a) and stabilizing (Fig. 4b).

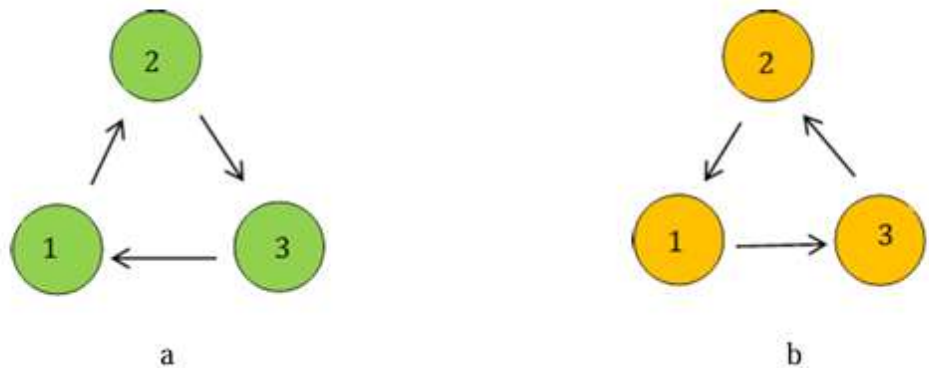

**Figure 4**: Schematic representation of self-organizing (a) and stabilizing (b) cycles in a system of three agents (adapted from Ulanowicz, 2009)

The trade-off between self-organization and historical organization determines the sign of the redundancy $R$ in Eq. 4.

Redundancy, as defined by Eqs. 1, 2 is a static measure, but real systems are not static. The system may be in one or more modes simultaneously. Assuming that the system is in a mode that involves smaller fluctuations around the average redundancy, the time evolution ща еру ыныеуь can be captured by a nonlinear evolutionary equation (Ivanova, Rzadkowski, 2024a):

$$4R_T - 2RR_X + R_{XXX} + C_1 = 0 \qquad (5)$$

Here $T$ refers to time and $X$ refers to phase space. Eq. 5 is a modification of Korteveg-de Vries (KdV) equation:

$$u_T - 6uu_X + u_{XXX} = 0$$

A distinctive feature of KdV equation is the existence of stable solitary wave solutions (solitons) of the form:

$$u = -\frac{1}{2}k^2 ch^{-2}\left[\frac{k}{2}(X - k^2 T)\right]$$

For a soliton chain, the ratio of soliton amplitudes to time shifts from the origin is constant. The total redundancy is the result of two competing sub-dynamics: synergy (indicated by the positive time series) and historical generation of variations (indicated by the negative time series).

## 4. Measurement

Data analysis is performed to detect patterns in the data and to find explanations for these patterns. According to Equation 5, redundancy can be expected to appear in (one or more) solitary wave trains whose amplitudes follow a linear trend. This suggests that the data are analyzed in terms of soliton solutions to the KdV equation in order to extract meaningful features of the data. An increase in negative redundancy indicates an increase in synergy (cohesive effort of parts of the system), which leads to an increase in the output reflected in the data. To extract relevant patterns from the data, one may choose to decompose the time-cumulative data into a series of logistic (S-shaped) curves. While cumulative data are represented as a sum of S-shaped curves, shifted along the time axis, differential (e.g. daily) data are represented as a sum of KdV solitons due to the correspondence of the derivative of the logistic curve to the KdV soliton.

The logistic function is the solution to the logistic equation:

$$x'(t) = \frac{s}{x_{sat}} x(x_{sat} - x) \qquad (6)$$

which has the form:

$$x(t) = \frac{x_{sat}}{1 + e^{-s(t - t_0)}} \qquad (7)$$

The derivative of the logistic function (6) is equal to:

$$x(t) = \frac{x_{sat}}{4} cosh^{-2}[s(t - t_0)] \qquad (8)$$

Differentiating the sum of logistic components yields a sum of hyperbolic functions, which can be analyzed in terms of solutions of the KdV equation.

According to the theory, self-organizing dynamics can be traced in systems of three or more heterogeneous agents. Each agent can include a number of elements, and "heterogeneous" means positional differentiation. This condition is met by a number of systems of different origins. Examples are financial, stock, energy, food and other markets. Markets include three differentiated groups of agents according to their position regarding work with market assets: those who prefer to buy, sell and those who refrain from buying or selling, waiting for more suitable conditions to enter the market.

The stock market time series can be represented as a sequence of discrete regions of finite length, including uptrends (periods with rising prices), downtrends (periods with falling prices), and periods of price consolidation. Each region in the time series is generated by the system with different parameters. An example of the soliton decomposition of daily Corn market data is shown in Figure 5. The data correspond to the period 2020.08.07 -2021.07.14.

Filtering the time series data in the solitary wave region produces a chain of consecutive solitary waves that indicate synergies generated within the system.

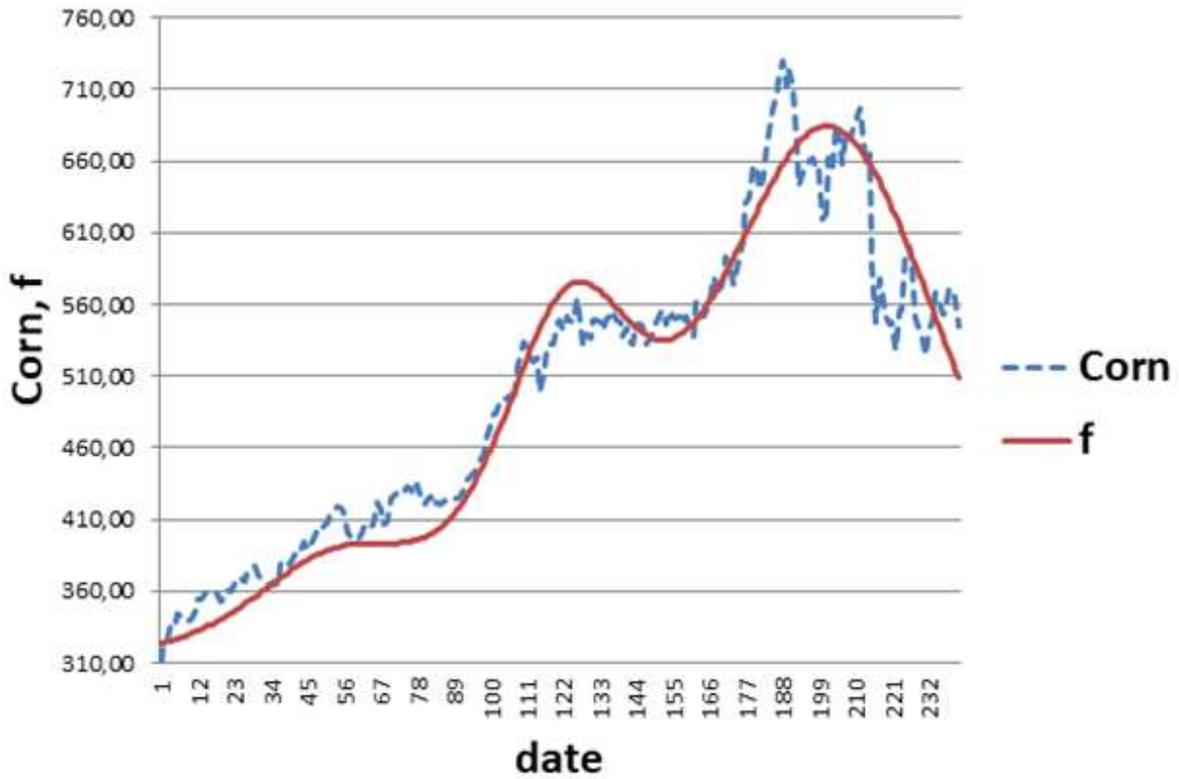

**Figure 5**: Decomposition of the Corn/USD data in a set of three solitons

The approximation parameters are given in Table 1.

**Table 1** Approximation parameters for Corn/USD pair

| β | $A_1$<br>$(k_1;\ T_1)$ | $A_2$<br>$(k_2;\ T_2)$ | $A_3$<br>$(k_3;\ T_3)$ |
|---|---|---|---|
| 310.75 | 71.75<br>(0.03; -54.16) | 208.21<br>(0.04; -122.4) | 370.57<br>(0.02; -201) |

Here $A_i$ are the amplitudes of solitons ($i$ - refers to the soliton number in the series), $k_i$ are the $k$ parameters, $T_i$ are the time spans, $\beta$ – is the vertical shift (introduced to equate the beginning of the first wave to zero). Ordinary least square fit $f = Bt + C$ allows the model to explain 94% of the variations (Fig. 5).

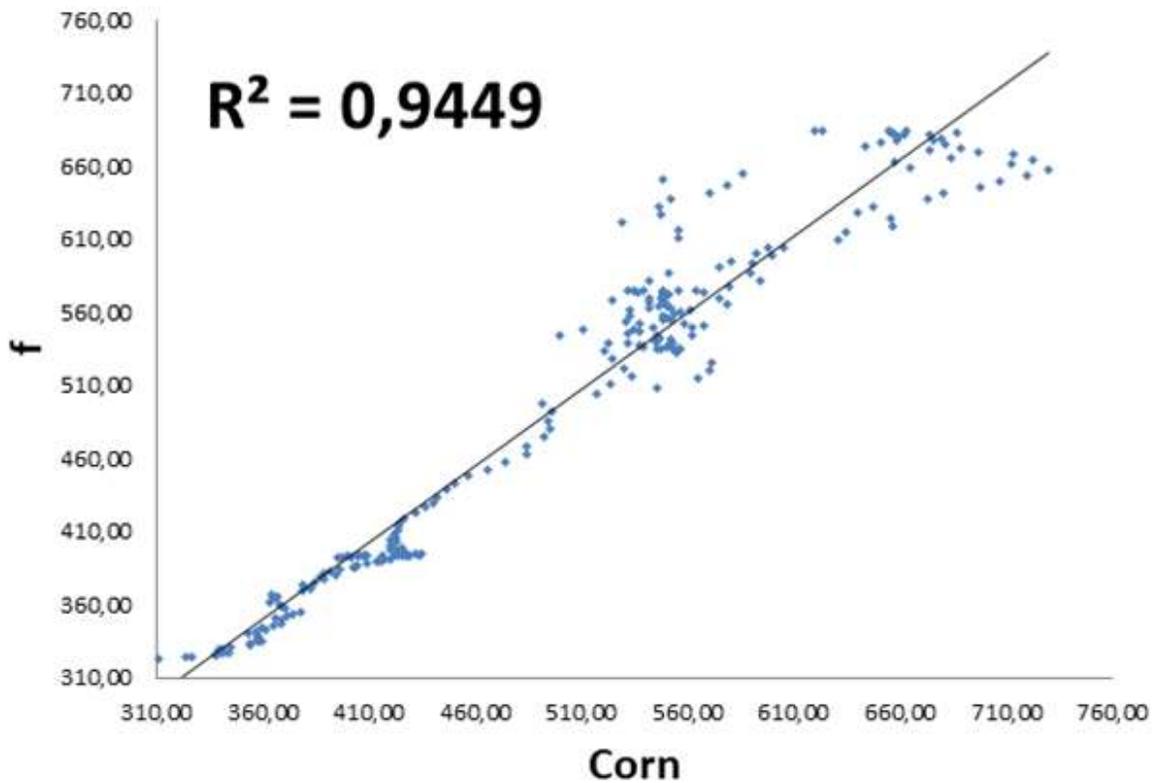

**Figure 5:** Scatter chart for Corn/USD values vs. model predicted values (*f*). Straight line is the OLS fit.

The parameters of the OLS are presented in Table 2.

**Table 2** OLS regression parameters (t-values are in parentheses)

| B | Constant | Observations | Adjusted $R^2$ |
|---|---|---|---|
| 0.95*<br>(7.26) | 52.39*<br>(64.03) | 241 | 0.94 |

*p<0.01

To eliminate the problem of spurious correlation between the asset price and the predicted curve model, one can perform a two-step Engel-Granger co-integration test (Engel & Granger, 1987) by regressing the asset price with the approximation function and an intercept, and then run an Augmented Dickey-Fuller (ADF) test for stationarity. The null hypothesis of a unit root can be rejected at the 1% level.

The logistic decomposition shows the resulting synergy generated in the system. But the resulting synergy is a trade-off between historical realization and self-organization, as indicated by the two terms in Eq. 4. A more detailed analysis should allow the contribution of each term to be assessed separately. This can be done using the wavelet transform or, more specifically, the logistic Continuous Wavelet Transform (CWT). The wavelet transform expands the time series data in terms of wavelet functions that are localized in both time and frequency. The logistic CWT decomposes the data set into an elementary form representing a sum of logistic functions. Like the Fourier transform, which converts a signal into a sum of harmonic functions representing the frequencies that are present in the function, the wavelet transform converts a function into a sum of wavelets. Wavelets are wave-like oscillations whose amplitudes differ from zero only on limited time intervals. A set of wavelets can be obtained from a parent wavelet by dilating and translating. Wavelets are more suitable than the Fourier transform for analyzing short signals. Another advantage of the wavelet transform over other methods is that it eliminates the problem of non-stationarity in the time series. The time series and any of its differences do not necessarily have to be stationary. The wavelet transform method can be used for both stationary and non-stationary series.

The result of the wavelet transform is presented as a wavelet scalogram. A scalogram is a two-dimensional representation of one-dimensional data. The X axis represents time, and the Y (scale) axis represents the result of the wavelet transform of the time series, corresponding to the value of the wavelet amplitude at time X. The analytical significance of such a graphical representation of the time series is that the Y axis represents the time resolution relative to the wavelet scale, which provides additional information about the dynamic properties of the signal.

Typically, time series are approximated by a sum of wavelets. If there are multiple overlapping waves occurring in the same time period, the waves with higher intensity may cause the waves with lower intensity to be invisible in the scalogram. We can remove the first wave with the highest intensity by subtracting it from the time series to make the waves with lower intensity visible. Figure 7 shows the wavelet scalegram of the CWT analysis for patent applications filed in the United States by residents for the period 1980-2021 (Fig. 6) before and after subtracting the largest wave.

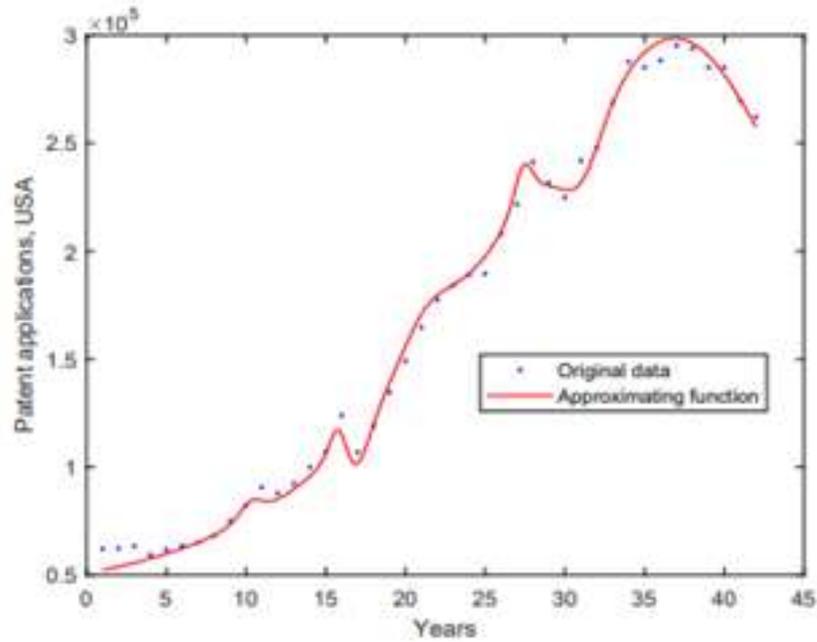

**Figure 6:** Patent applications filed in the United States by residents of the country for the period 1980-2021, the dotted line is statistical data, the solid line is the approximation function (From Ivanova and Rzadkovsky, 2024b)

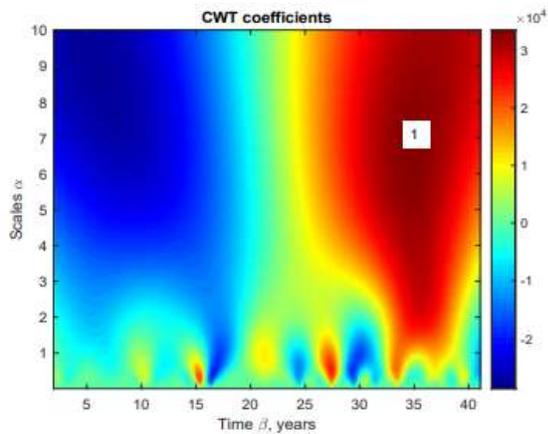

(a) Wave no 1

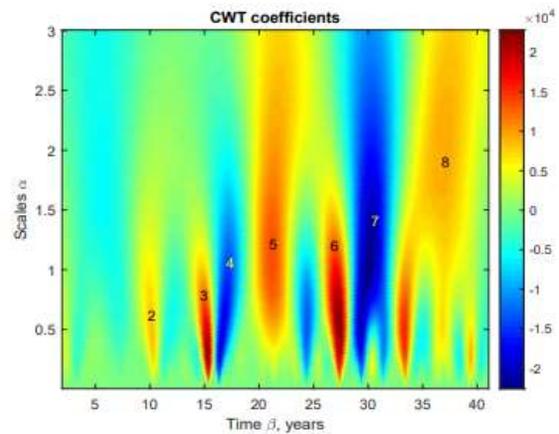

(b) Waves after removing the wave no 1

The results shown in Fig. 7 are presented in a more visual form in Fig. 8

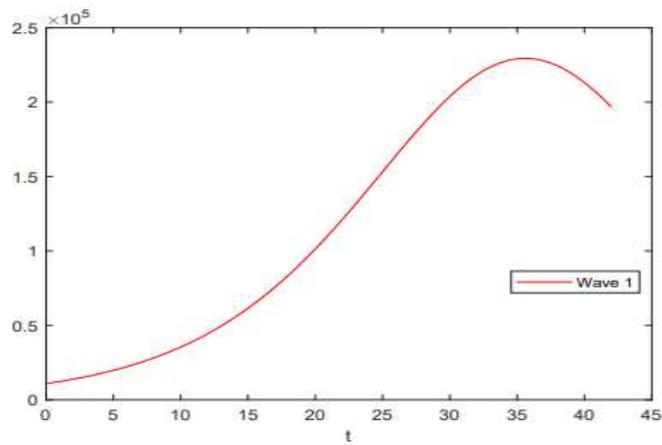

a)

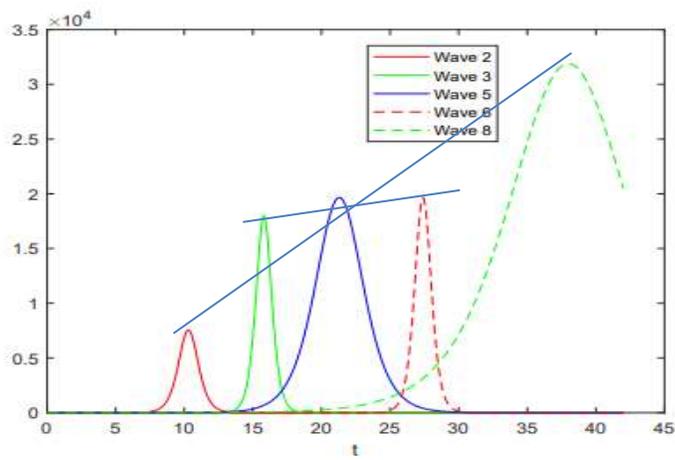

b)

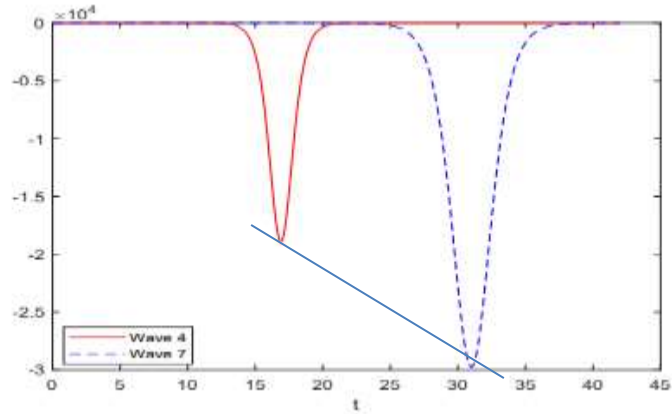

c)

**Figure 8:** Solitary wave decomposition of a yearly data on patents applications filed in USA by country residents for the period 1980-2021, as shown in Fig. 6 (From Ivanova and Rzadkovsky, 2024a)

The waves in Fig. 8 (b and c) follow the redundancy distribution predicted by the TH model. The wave peaks in each wave train form a linear trend and can be interpreted in terms of different coexisting modes. Fig. 8(c) shows two sequences of peaks that can be interpreted as different modes coexisting at the system level. The linear trend formed models the general direction in which the data is moving and can be useful for finding change points in the time series.

The advantage of using logistic CWT analysis instead of logistic decomposition is that CWT can reveal the complex linear structure of the soliton trend, even if it is difficult to guess. CWT can also separately account for positive and negative trend components and different modes.

## 5. TH model – biological turn

The Triple Helix (TH) metaphor is also of interest from a biological perspective because it is homologous (of the same origin) with embryologic development, culminating in physiologic control of the organism. And beyond that, it forms the basis for injury repair, enabling the organism to pass its genes to the next generation.

Organisms without a mesoderm remain relatively primitive, like jellyfish, though their bauplan is adequate for their environment. On the other hand, those organisms with a mesoderm are much more dynamically suited in adaptating to an ever-changing environment. It is the mesodermal-endodermal and mesodermal-ectodermal interactions during embryogenesis that form the basis for progressively more complex organisms, having organs and organ systems to accommodate their environmental challenges. The means for doing so has been described by Gould and Vrba (1982) as exaptations, or referring back to earlier adaptations to solve current problems. But it should be emphasized that the progressive complexification of physiology is in service to the body plan acting as the 'history' of the organism as the means of determining environmental change in order to effectively adapt, the core simplicity of the unicell remaining intact genetically (Torday, 2016). And when injured, these 'triplex' organisms recapitulate their developmental motifs to re-establish homeostatic control of their cellular environment. And perhaps the reason for homology between TH and biology is the homology between the atom and the sell (Torday, 2022). In both cases, the construct is based on both deterministic and probabilistic principles.

One can find the homology between embryologic development and the TH model of reflexive communications (Table 3).

**Table 3** A homology between embryologic development and TH model of reflexive communications

| cells | agents |
|---|---|
| loss of cellular homeostasis, followed by the production of Radical Oxygen Species | an increase of system's entropy in accordance with the Second law of thermodynamics |
| physiological local consciousness | communication between positionally identical agents<br>(sent and received messages are processed with the same set of communication codes) |
| physiological non-local consciousness | communication between positionally different agents<br>(sent and received messages are processed with different sets of communication codes) |
| local | relational (as in a network) |
| non-local | positional (different positions occupied in a vector space of communication codes) |
| structural and functional 'remodeling' | realization of additional options |
| disruptions are resolved through the re-establishment of cellular energetic homeostasis | The system evolves through self-organization as indicated by negative redundancy values |
| reference to both physiological local and non-local consciousness | Redundancy (R) as a result of two competing dynamics: historical realization (P) and self-organization (Q)<br>$$R = P - Q$$ |

The 'economy' is most effective when it aligns with our physiology. Witness Jared Diamond's book "Collapse", which is predicated on humans most attuned with their natural environment being those that have been most successful, i.e. through their physiology aligning with Nature (Diamond, 2011). As an example of how our physiology is attuned with the economy, the Bayh-Dole Act of 1980 undermined academic biomedical research, putting it on par with industrial biomedical research. In so doing, it eliminated the 'freedom' of thought in research interfacing with its ever-changing environment, instead holding the latter to a monetary standard, undercutting how and why academic biomedical research seeks out novel ideas in deference to technical advances. That phenomenon echoes Eisenhower's "beware the military-industrial complex' voiced upon his departure from the Presidency of the United States (Eisenhower, 1961). That warning is more important than ever. The 'three legs' of the TH 'stool' are sustainable, whereas the two-legged stool is not.

In fact, the Bayh-Dole Act can be seen as a mechanism that transforms the TH system into a Double Helix (DH) system, thereby fundamentally undermining its performance. The negative impact of this law may weaken over time, as the national innovation system eventually evolves from a TH system to a Quadruple Helix (QH) system by attracting additional actors, thereby restoring the broken TH symmetry (Ivanova, 2014).

## 6. Discussion

The quantitative theory of meaning is closely related to the TH model. The TH model, being the simplest representation of a nonlinear self-organizing system, can be used as a building block for higher-order systems. The TH metaphor invites extension to quadruple, quintuple and higher order helices. Carayannis and Campbell (2009, 2010) proposed quadruple helix system consisting of media- and culture-based societies and quintuple helix systems that include the natural environment of society and the economy. However, these generalizations in their extension do not support the rotational symmetry inherent in TH. One of the authors of this article proposed to use "media" as the fourth helix to extend the TH model with respect to rotational symmetry, so that the evolution of the model is described as a rotation of the system representation vector in the four-dimensional university-industry-government-media space (Ivanova, 2014).

Leydesdorff mentioned that "… *an N-tuple or an alphabet of (20+) helices can be envisioned. However, in scholarly discourse and for methodological reasons, one may wish to extend models step-by-step and as needed to gain explanatory power* … " (Leydesdorff, 2012, at. p. 32).

TH reveals all the main features of nonlinearity and self-organization inherent in higher-order systems. TH synergy serves as an indicator of system performance and stability, which can also be attributed to the complexity indicator.

Measuring system complexity has long attracted the attention of researchers and practitioners. In economic research, various complexity indicators have been proposed to assess the relative economic complexity in a set of countries in terms of countries and products exported by countries (e.g. Hidalgo and Hausmann, 2009; Tacchella *et al*. 2013). Exported products are assumed to reflect the available diversified and non-ubiquitous production capabilities that a country possesses. The country-product network can be extended by adding a technological (patent) dimension (Ivanova, Smorodinskaya, Leydesdorff, 2019). A distinctive feature of complexity indicators is that they can be used to predict future growth of GDP per capita.

The relationship between complexity and sustainability in its economic aspect implies that the sustainable economic development of a country is ensured by a "complex" economy with diversified capabilities. These capabilities also add economic competitive advantages and the economy's ability to adapt to the changing socio-economic environment. But is there a relationship between these measures and TH synergy? A study of the synergy distribution among 31 Chinese territorial districts and the corresponding complexity distribution shows a significant correlation between synergy and complexity measures (Ivanova, 2022). That is, information (synergy) and production (complexity) measures are interrelated.

An analysis of patent application data for four countries - Switzerland, the United States, Germany and Brazil, filed between 1980 and 2021 by residents of the countries, using the procedure of decomposition of longitudinal patent data with the logistic continuous wavelet transform reveals a hidden trend structure that follows solitary wave patterns consisting of positive and negative series of KdV solitons (Ivanova, Zhadkovsky, 2024a). This structure indicates the efficiency of the innovation system in terms of patent generation. Positive series can be attributed to the synergy generated within the system, and negative series indicate an increase in entropy due to historical evolution. Several modes can operate simultaneously. Thus, the innovation system can be assessed in terms of systemness, where synergy plays a leading role in the sustainable development of the system.

The method for analyzing information processing and meaning generation in social systems is universal in nature and can be extended to different types of social systems. Application to financial markets – in particular, the EUR/USD case study – reveals the same underlying trend structure (Ivanova, Zhadkovsky, and Leydesdorff, 2024b).

The growing understanding that investors assess market situations from different perspectives, which forms the basis for further actions, has led to the emergence of a new paradigm for understanding financial markets – social finance (Hirshleifer, 2020; Akçay, Hirshleifer, 2021). Social finance postulates that accepted cultural traits – beliefs, strategies, and folk economic models – are inherent in larger groups of investors and provide informational biases for transmission among investor groups, which are the driving forces behind investor decisions. In this vein, different cultural traits, which are subject to different biases in judgments and decisions, act as different sets of communication codes, generating meaning from information available in the market.

With regard to the TH metaphor, we can also see a deep homology between TH and evolution, in particular the role of the mesoderm in gastrulation as the introduction of a third germ layer into the cell (Torday, 2021). The three layers are the precursors of all embryonic tissues.

Network of relations and network of correlations can be considered as implicate and explicate orders. Implicate and explicate orders accentuate the primacy of structure over structural elements.  In this respect meanings, generated via communications (which can't be observed but only inferred) and actions (observed phenomena) appear differently as instantiations of implicate and explicate orders.

In the presented approach, the complexity of the system arises as a consequence of the topology of TH. And the simplicity of TH simplifies the understanding of the mechanisms that govern the dynamics of a complex system. The fractal structure of TH can be considered as an intrinsic property of complex systems. The hologrammatic principle states that the part is identical to the whole. In this respect, the hologram resembles a fractal, which is also constructed as a self-referential structure. A similar fractal structure can be found in inherently nonlinear systems such as TH, where communications generate communications.

Unlike Shannon's Communication Theory, the Theory of Meaning imposes restrictions on the complex systems analyzed. To be nonlinear, they must include a TH-like structure which entails certain dynamics. This specific dynamics also invites a method for analyzing complex systems, such as the logistic CWT. In this respect, the Theory of Meaning resembles string theory in particle physics, which is not background-independent and defines a fixed reference geometry for space-time. There is another interesting question that requires further research. If TH can generate complexity, can complex systems exhibit a TH-like structure? In other words, to what extent can the TH structure be considered a fundamental feature of complex systems.

## 7.  Conclusion

Information theory, being an abstract free content theory, can be a useful tool for studying complex systems. While general information theory does not imply any structure in the systems under study, adding the dimension of meaning changes the situation. The theory of meaning appears to be closely related to the TH model, which initially emerged as a model of the national innovation system of university-industry-government relations. The mutual co-evolution of these two approaches led to a fairly general concept of a system governed by reflexive information communication. Thus, the structure and dynamics of the system turned out to be interrelated.

 Here we briefly summarize the main results related to the TH metaphor.

1. The evolution of the system is largely ensured by the exchange of information within the system.

2. The interaction between the positionally differentiated agents that make up the system ensures proliferation of additional, not yet realized options that may be more important for the future development of the system than those already realized. The number of additional options is of decisive importance for the sustainability of the system.

3. The nonlinear dynamics of the system is ensured by the corresponding structure of the system and the positional differentiation of the agents included in it, where the number of agents must be three or more.

4. A system with three or more agents can generate a fractal structure. This fractal structure can be empirically observed in star-shaped networks of relationships.

5. The system produces both positive and negative entropy during its evolution. Positive entropy is generated with the arrow of time as a result of historical development, and negative entropy arises from self-organization and is generated against the arrow of time. Negative entropy can be considered as a synergy of agent interaction.

6. When analyzing empirical data of complex systems from the point of view of information exchange between heterogeneous agents, the most effective method of analysis appears to be the logistic CWT. The method allows not only to select a specific type of wavelet, but also to give conceptual explanations for the results obtained.

The scope of the TH metaphor is not limited to innovation studies. Without loss of generality, it can be applied as a universal concept in completely different areas – from knowledge generation to financial research, the spread of infectious diseases, rumors propagation, etc. Moreover, TH, as the simplest nonlinear system with the main features inherent in higher-order systems, can be used as a building block for these higher-order systems, since triadic closure can be seen as the basic mechanism of community generation in complex networks. Higher-order systems such as quadruplet, quintet, sextet, etc. can be reduced to triads (Bianconi *et al*., 2014; Freeman, 1996).

We tend to see complexity from the manifestation of its effects rather than from its primary causes and without taking into account the topology of the system. Complex systems usually consist of many interacting elements. But this number of elements can be structured with respect to their positional differentiation, which simplifies system analysis. In this respect, considering the complexity of a system as a consequence of the combination of basic elementary components of the system, such as TH, can be more constructive for its study and understanding.

## Acknowledgments

The Authors are grateful to Mark Johnson for his valuable comments on the paper topic.

## References

Abramson, N. (1963). *Information Theory and Coding*. New York, etc.: McGraw-Hill.

Akçay, E. and Hirshleifer, D. (2021), Social finance as cultural evolution, transmission bias, and market dynamics, *Proceedings of the National Academy of Science, 118*(26), 2015568118- . https://www.pnas.org/doi/pdf/10.1073/pnas.2015568118

Albert, R., Jeong, H., & Barabási, A. L. (2000). Error and attack tolerance of complex networks. *Nature*, *406*(6794), 378-382.

Bar-Yam, Y. (2004). *Making Things Work: Solving Complex Problems in a Complex World*. Knowledge Press.

Bianconi, G.,Darst,R.K.,Iacovacci,J.,and Fortunato,S. (2014). Triadic closure as a basic generating mechanism of communities in complex networks. Physical Review E, *90*(4), 042806.

Biggiero, L. and Angelini, P. P. (2015). Hunting scale-free properties in R&D collaboration networks: Self-organization, power-law and policy issues in the European aerospace research area, *Technological Forecasting & Social Change*, *94*, 21–43.

Carayannis, E. G., & Campbell, D. F. J. (2009). 'Mode 3' and 'Quadruple Helix': toward a 21st century fractal innovation ecosystem. *International Journal of Technology Management, 46*(3), 201–234.

Carayannis, E. G., & Campbell, D. F. J. (2010). Triple Helix, Quadruple Helix and Quintuple Helix and how do knowledge, innovation, and environment relate to each other? *International Journal of Social Ecology and Sustainable Development, 1*(1), 41–69.

Cover, T. M., & Thomas, J. A. (2006). *Elements of information theory*. John Wiley & Sons.

Diamond, J. (2011). *Collapse: How Societies Choose to Fail or Survive*, Penguin Books

Eisenhower, D. (1961), "President Dwight D. Eisenhower's Farewell Address". https://www.archives.gov/milestone-documents/president-dwight-d-eisenhowers-farewell-address. Accessed Dec. 10, 2024.

Engle, R., Granger, C. (1987). Co-integration and error correction: Representation, estimation and testing. *Econometrica*. *55* (2), 251–276.

Etzkowitz, H. and Leydesdorff, L. (1995), "The Triple Helix - University-Industry-Government Relations: A Laboratory for Knowledge-Based Economic Development", *EASST Review, 14*(1), pp. 14-19.

Etzkowitz, H. and Leydesdorff, L. (1998), The endless transition: A "triple helix" of university – industry – government relations, *Minerva, 36*, 203-208.

Etzkowitz, H., and Leydesdorff, L. (2000). The Dynamics of Innovation: from National Systems and "mode 2" to a Triple Helix of university-industry-government relations, *Res. Policy, 29* (2), 109-123.

Etzkowitz, H., Ranga, M. (2012). "Spaces": A triple helix governance strategy for regional innovation. In: A. Rickne, S. Laestadius & H. Etzkowitz (Eds.), *Innovation Governance in 39 an Open Economy: Shaping Regional Nodes in a Globalized World*, Milton Park, UK: Routledge.

Freeman, L. C. (1996). Cliques, Galois lattices, and the structure of human social groups. *Social Networks, 18*(3), 173-187.

Gould, S.J., Vrba, E.S. (1982). Exaptation- a missing term in science of form. *Paleobiology,* 8, 4-15.

Hidalgo C. A., Hausmann R. (2009). "The Building Blocks of Economic Complexity", *Proceedings of the National Academy of Sciences of the United States of America*. *106*(26), 10570–10575.

Hirshleifer, D. (2020). "Presidential address: Social transmission bias in economics and finance", *J. Finance*, 75, 1779–1831.

Ivanova, I. (2014). Quadruple Helix Systems and Symmetry: a Step Towards Helix Innovation System Classification, *Journal of the Knowledge Economy 5*(2), 357-369. https://doi.org/10.1007/s13132-014-0201-z

Ivanova, I. (2022). "The Relation between Complexity and Synergy in the case of China: Different ways of predicting GDP in a complex and adaptive system", *Quality &Quantity*, *56*, 195-215. doi: 10.1007/s11135-021-01118-6

Ivanova, I. and Leydesdorff, L. (2014). Rotational Symmetry and the Transformation of Innovation Systems in a Triple Helix of University-Industry-Government Relations, *Technological Forecasting and Social Change*, *86*, 143-156; http://dx.doi.org/10.1016/j.techfore.2013.08.022

Ivanova, I., Rzadkowski, G. (2024a). Triple Helix synergy and patent dynamics. Cross country comparison. *Quality & Quantity* (in preparation). https://arxiv.org/abs/2406.15780 Accessed Dec. 10, 2024.

Ivanova, I., Rzadkowski, G. and Leydesdorff, L. (2024b). Quantitative Theory of Meaning. Application to Financial Markets. EUR/USD case study. (in preparation). https://arxiv.org/abs/2410.06476 Accessed Dec. 10, 2024.

Ivanova I, Smorodinskaya, N., and Leydesdorff, L. (2019). On measuring Complexity in a PostIndustrial Economy: The Ecosystem's Approach. *Quality & Quantity 54*(1), 197-212. doi: 10.1007/s11135-019-00844-2

James, R., Ellison, C., and Crutchfield, J. (2011). Anatomy of a bit: Information in a time series observation, *Chaos: An Interdisciplinary Journal of Nonlinear Science, 21*, 037109

Kuehn, E. (2023). The information ecosystem concept in information literacy: A theoretical approach and definition. *JASIST*, *4* (3), 434-443.

Leydesdorff, L. (2010). The Knowledge-Based Economy and the Triple Helix Model. *Annual Review of Information Science and Technology*, *44*, 367-417

Leydesdorff, L. (2012). The Triple Helix, Quadruple Helix …, and an N-tuple of Helices: explanatory models for analyzing the knowledge-based economy. *Journal of the Knowledge Economy, 3*(1), 25–35.

Leydesdorff, L. and Ivanova, I. (2014). Mutual Redundancies in Inter-human Communication Systems: Steps Towards a Calculus of Processing Meaning, *Journal of the Association for Information Science and Technology 65*(2), 386-399. doi: 10.1002/asi.22973

Leydesdorff, L., Petersen,A., and Ivanova, I. (2017). Self-Organization of Meaning and the Reflexive Communication of Information. *Social Science Information 56*(1), 4-27. doi: 10.1177/0539018416675074

Li, M., & Vitányi, P. (2008). *An introduction to Kolmogorov complexity and its applications*. Springer Science & Business Media.

McGill, W. J. (1954). Multivariate information transmission. *Psychometrika, 19*(2), 97-116.

Monleón-Getino, T., Rodríguez-Casado, C., and Verde, P. (2019). Shannon Entropy Ratio, a Bayesian Biodiversity Index Used in the Uncertainty Mixtures of Metagenomic Populations. *Journal of Advanced Statistics*, 4(4), 23-34.

Shannon, C. E. (1948). A Mathematical Theory of Communication. *Bell System Technical Journal, 27*, 379-423 and 623-656.

Shaw, D., Allen, T. (2018). Studying innovation ecosystems using ecology theory. *Technol. Forecast. Soc. Chang. 136*, 88–10.

Sporns, O. (2010). *Networks of the brain*. MIT press.

Steuer, R., Kurths, J., Daub, C. O., Weise, J., & Selbig, J. (2002). The mutual information: detecting and evaluating dependencies between variables. *Bioinformatics*, *18*(suppl_2), S231-S240.

Strong, S. P., Koberle, R., de Ruyter van Steveninck, R. R., & Bialek, W. (1998). Entropy and information in neural spike trains, *Physical review letters*, *80*(1), 197-200.

Sun, Y., & Negishi, M. (2010). Measuring the relationships among university, industry and other sectors in Japan's national innovation system: A comparison of new approaches with mutual information indicators. *Scientometrics, 82*(3), 677–685.

Tacchella, A., Cristelli, M., Caldarelli, G., Gabrielli, A., & Pietronero, L. (2013). Economic complexity: conceptual grounding of a new metrics for global competitiveness. *Journal of Economic Dynamic and Control*, *37*(8), 1683-1691.

Torday, J.S. (2016). Life Is Simple-Biologic Complexity Is an Epiphenomenon. *Biology* (Basel) *5*, 17.

Torday, J. (2021). Cellular evolution of language. *Prog Biophys Mol Biol.,167*, 140-146.

Torday, J. (2022). Quantum Mechanics, Cell-Cell Signaling, and Evolution. Academic Press, New York

Trigona, R. and Cianci, E. (2014) Glossary of Complexity, *World Futures: The Journal of New Paradigm Research, 70*(5-6), 370-375.

Ulanowicz, R. E. (2009). The dual nature of ecosystem dynamics. *Ecological Modelling, 220*(16), 1886–1892.

Varley, T. F. (2023). Information Theory for Complex Systems Scientists: What, Why, & How? https://arxiv.org/pdf/2304.12482 Accessed Dec. 10, 2024.

Yeung, R. W. (2008). *Information Theory and Network Coding*. New York, NY: Springer.